\documentclass[12pt]{iopart}
\usepackage{setstack}

\begin{document}

\title{Self-energy effects in the Polchinski and Wick-ordered
renormalization-group approaches}
\author{A. Katanin \\
Institute of Metal Physics, 620041, Ekaterinburg, Russia}

\begin{abstract}
I discuss functional renormalization group (fRG) schemes, which allow for
non-perturbative treatment of the self-energy effects and do not rely on the
one-particle irreducible functional. In particular, I consider Polchinskii
or Wick-ordered schemes with amputation of full (instead of bare) Green
functions, as well as more general schemes, and eastablish their relation to
the `dynamical adjustment propagator' scheme by M. Salmhofer [Ann. der Phys.
16, 171 (2007)]. While in the Polchinski scheme the amputation of full
(instead of bare) Green functions improves treatment of the self-energy
effects, the structure of the corresponding equations is not suitable to
treat strong-coupling problems; it is not also evident, how the mean-field
(MF) solution of these problems is recovered in this scheme. For Wick
ordered scheme, excluding fully or partly tadpole diagrams one can obtain
forms of fRG hierarchy, which are suitable to treat strong-coupling
problems. In particular, I emphasize usefullness of the schemes, which are
local in cutoff parameter, and compare them to the one-particle irreducible
approach.
\end{abstract}

\maketitle

\section{Introduction}

Applications of functional renormalization group (fRG) approach \cite%
{Polchinski,SalmhoferBook,Wetterich} to problems of condensed matter and
high energy physics have received substantial progress recently\cite{Review}%
. However, correct treatment of the self-energy effects remains one of the
most difficult points of many fRG schemes.

Currently, the one-particle irreducible (1PI) scheme\cite%
{Wetterich,SalmhoferRev,Honer,Review,KataninKampf,HonerU,Katanin1} is mostly
used for treatment of the self-energy effects within fRG. Simple truncations
within this scheme, however, do not allow to fulfill Ward identities and
sometimes do not reproduce even the results of the mean-field approach, for
which correct treatment of the self-energy effects is crucial. The
truncation of 1PI equations of Ref. \cite{Katanin} allowed to reproduce the
mean-field results and improve the results of more sophisticated approaches
to fulfill Ward identities better. This approach found its applications in
the treatment of weakly- and moderately interacting single-impurity problems%
\cite{Pruschke}, Hubbard model in the symmetry broken phase\cite{Metzner},
and, more recently, two- and three dimensional Heisenberg model\cite{Woelfle}%
.

At the same time, 1PI approach has severe problems in describing the strong
coupling regime of many-body systems. The problem of application of this
approach in this case is mainly related to the one-loop structure of 1PI
hierarchy of fRG equations. Although higher-loop contributions can be
obtained by reinserting of the one-loop contributions from higher- to
lower-order vertices, this procedure is usually spoiled by truncations of
the hierarchy and the projection schemes used in approximate solutions,
which do not allow getting correct results for non-ladder diagrams at a
certain loop order\cite{KataninTwoLoop}. In particular, the truncation,
which neglects 8-point vertices and accounts for the full frequency- and
momentum dependence of the 4-point and 6-point vertices (which is already a
very complicated problem), yields a correct result for two-loop diagrams
only \cite{KataninTwoLoop}.

On the other hand, some non-1PI schemes, in particular Wick-ordered scheme%
\cite{SalmhoferBook}, already in the original formulation explicitly include
many-loop contributions and can, in principle, improve treatment of the
strong-coupling regime. The Wick-ordered scheme in its original formulation
is however not suited to treat self-energy effects in the non-perturbative
way. The change from amputation of the bare to interacting Green functions,
proposed in Ref. \cite{SalmhoferBook}, is not sufficient for such a
treatment, since the function, with respect to which the Wick ordering is
performed, should be also adjusted appropriately. A way for non-perturbative
treatment of the self-energy effects was considered in the `dynamical
adjustment' scheme of Ref. \cite{Salmhoferadj}. In the present paper we
propose somewhat different approach for non-perturbative treatment of the
self-energy effects and consider in detail both, local schemes, which do not
remove fully tadpole terms from the RG hierarchy, as well as non-local
schemes, removing fully tadpole terms. We compare the approaches of the
paper to the dynamical adjusting scheme of Ref. \cite{Salmhoferadj} and the
1PI approach, as well as discuss the results of the application of these
methods to the toy model.

\section{Self-energy effects in the Polchinskii scheme}

We consider the model described by an action%
\begin{equation}
\mathcal{S}_{\Lambda }[\overline{\psi },\psi ]=\mathcal{S}_{0}^{\Lambda }[%
\overline{\psi },\psi ]+\mathcal{V[}\overline{\psi },\psi ],  \label{S}
\end{equation}%
where $\overline{\psi },\psi $ are the bosonic or fermionic (Grassman)
fields, $\mathcal{V}$ is an interaction and%
\begin{equation}
\mathcal{S}_{0}^{\Lambda }[\overline{\psi },\psi ]=\int d^{d}x\int d\tau
\overline{\psi }(C_{0,\Lambda })^{-1}\psi  \nonumber
\end{equation}%
is quadratic in fields. $C_{0,\Lambda }$ is the cut bare propagator, e.g.
having the Fourier transform $C_{0,\Lambda }=\chi _{>,\Lambda }/(i\omega
_{n}-\varepsilon _{\mathbf{k}}),$ where $i\omega _{n}$ are bosonic or
fermionic Matsubara frequencies, $\varepsilon _{\mathbf{k}}$ is the
dispersion, and $\chi _{>,\Lambda }$ is the cutoff function, which cuts off
the the low-energy modes, e.g. $\chi _{>,\Lambda _{0}}=0$ and $\chi _{>,0}=1$%
.

The standard procedure of derivation of fRG equations relies on the
differentiating of generating functional for connected vertices, amputated
with the non-interacting Green functions,%
\begin{equation}
\mathcal{V}_{\Lambda }[\overline{\psi },\psi ]=-\ln \int D[\overline{\chi }%
,\chi ]e^{-\mathcal{S}_{0}^{\Lambda }[\overline{\chi },\chi ]-\mathcal{V}[%
\overline{\psi }+\overline{\chi },\psi +\chi ]}
\end{equation}%
This procedure can be supplemented by a consequent reamputation of external
legs, changing their amputation from non-interacting to that with
interacting Green functions.

For the purposes of the present paper, we consider somewhat more general
procedure, which allows to obtain the renormalization-group equations
accounting for the self-energy effects with further possible
generalizations. To this end, we introduce the counterterm $\overline{\Sigma
}_{\Lambda }\overline{\psi }\psi $ into the bare action and replace $%
(C_{0,\Lambda })^{-1}\rightarrow C_{\Lambda }^{-1}$:%
\begin{equation}
\mathcal{S}_{\Lambda }[\overline{\psi },\psi ]=\overline{\psi }C_{\Lambda
}^{-1}\psi +\overline{\psi }\ \overline{\Sigma }_{\Lambda }\psi +\mathcal{V[}%
\overline{\psi },\psi ]
\end{equation}%
Although for the choice
\begin{equation}
C_{\Lambda }^{-1}=(C_{0,\Lambda })^{-1}-\overline{\Sigma }_{\Lambda }
\label{CL}
\end{equation}%
the counterterm is cancelled, we will not in general assume validity of
equation (\ref{CL}) during the flow, requiring its fulfillment only in the
end of the flow. We also treat $\overline{\Sigma }_{\Lambda }$ as a part of
the interaction, such that amputation of the vertices by the functions $%
C_{\Lambda }$ is achieved naturally in this way, since the non-interacting
term now contains $C_{\Lambda }^{-1}$. Introducing the effective interaction
in the standard way%
\begin{equation}
e^{-\mathcal{V}_{\Lambda }[\overline{\eta },\eta ]}=\int D[\psi ,\overline{%
\psi }]e^{-\overline{\psi }(C_{\Lambda }^{-1})\psi -\mathcal{V}[\overline{%
\psi }+\overline{\eta },\psi +\eta ]+\overline{\Sigma }_{\Lambda }(\overline{%
\psi }+\overline{\eta })(\psi +\eta )}
\end{equation}%
and performing the same steps, as for deriving Polchinski equation, we obtain%
\begin{equation}
\partial _{\Lambda }\mathcal{V}_{\Lambda }=\Delta _{\dot{C}_{\Lambda }}%
\mathcal{V}_{\Lambda }-\Delta _{\dot{C}_{\Lambda }}^{12}\mathcal{V}_{\Lambda
}^{(1)}\mathcal{V}_{\Lambda }^{(2)}+\dot{\overline{\Sigma }}_{\Lambda }\frac{%
\delta \mathcal{V}_{\Lambda }}{\delta \overline{\Sigma }_{\Lambda }}
\end{equation}%
where $\Delta _{\dot{C}_{\Lambda }}\mathcal{V}_{\Lambda }=\mathrm{Tr}(\dot{C}%
_{\Lambda }\delta ^{2}\mathcal{V}_{\Lambda }/\delta ^{2}\Psi )$, $\Delta _{%
\dot{C}_{\Lambda }}^{12}\mathcal{V}_{\Lambda }^{(1)}\mathcal{V}_{\Lambda
}^{(2)}=\left( \frac{\delta \mathcal{V}_{\Lambda }}{\delta \Psi },\dot{C}%
_{\Lambda }\frac{\delta \mathcal{V}_{\Lambda }}{\delta \Psi }\right) ,$ $%
\Psi =(\psi ,\overline{\psi })$, $\partial _{\Lambda }$ and dot stand for
the derivative over $\Lambda $. Expressing the variational derivative $%
(\delta \mathcal{V}_{\Lambda }/\delta \overline{\Sigma }_{\Lambda })$
through variational derivatives of $\mathcal{V}_{\Lambda }$ over fields, we
obtain%
\begin{eqnarray}
\partial _{\Lambda }(\mathcal{V}_{\Lambda }-\overline{\Sigma }_{\Lambda }%
\overline{\psi }\psi ) &=&\Delta _{\dot{C}_{\Lambda }-\dot{\overline{\Sigma }%
}_{\Lambda }C^{2}}\mathcal{V}_{\Lambda }-\Delta _{\dot{C}_{\Lambda }-\dot{%
\overline{\Sigma }}_{\Lambda }C^{2}\Lambda }^{12}\mathcal{V}_{\Lambda }^{(1)}%
\mathcal{V}_{\Lambda }^{(2)}  \nonumber \\
&&-\dot{\overline{\Sigma }}_{\Lambda }C\left( \overline{\psi }\frac{\delta
V_{\Lambda }}{\delta \overline{\psi }}+\psi \frac{\delta V_{\Lambda }}{%
\delta \psi }\right)  \label{Eq}
\end{eqnarray}%
Choosing $\overline{\Sigma }_{\Lambda }=0$, we arrive at the hierarchy of
fRG equations in the Polchinski scheme\cite{Polchinski}%
\[
\partial _{\Lambda }\mathcal{V}_{\Lambda }=\Delta _{\dot{C}_{0,\Lambda }}%
\mathcal{V}_{\Lambda }-\Delta _{\dot{C}_{0,\Lambda }}^{12}\mathcal{V}%
_{\Lambda }^{(1)}\mathcal{V}_{\Lambda }^{(2)}
\]%
Considering expansion of $\mathcal{V}_{\Lambda }$ in fields%
\begin{equation}
\mathcal{V}_{\Lambda }=\sum\limits_{m}V_{m}\Psi (x_{1})....\Psi (x_{m}),
\end{equation}%
this hierarchy reads%
\begin{eqnarray}
\dot{V}_{2} &=&-V_{2}\dot{C}_{0,\Lambda }V_{2}+V_{4}\circ \dot{C}_{0,\Lambda
},  \nonumber \\
\dot{V}_{4} &=&-4V_{2}\dot{C}_{0,\Lambda }V_{4}+V_{6}\circ \dot{C}%
_{0,\Lambda },  \nonumber \\
\dot{V}_{6} &=&-6V_{2}\dot{C}_{0,\Lambda }V_{6}+V_{4}\dot{C}_{0,\Lambda
}V_{4}+V_{8}\circ \dot{C}_{0,\Lambda },  \nonumber \\
&&...,  \label{Polch}
\end{eqnarray}%
$\circ $ denotes convolution with respect to internal momenta and
frequencies. On the other hand, for the choice $V_{2}=0$ (\ref{Eq}) yields
equations for the self-energy $\Sigma _{\Lambda }=\overline{\Sigma }%
_{\Lambda }$ and fully amputated vertex functions, which are denoted in the
following as $H_{m}:$
\begin{eqnarray}
\dot{\Sigma}_{\Lambda } &=&-H_{4}\circ (\dot{C}_{\Lambda }-\dot{\Sigma}%
_{\Lambda }C^{2}),  \label{Mod} \\
\dot{H}_{4} &=&H_{6}\circ (\dot{C}_{\Lambda }-\dot{\Sigma}_{\Lambda
}C^{2})-4H_{4}C_{\Lambda }\dot{\Sigma}_{\Lambda },  \nonumber \\
\dot{H}_{6} &=&H_{8}\circ (\dot{C}_{\Lambda }-\dot{\Sigma}_{\Lambda
}C^{2})+H_{4}(\dot{C}_{\Lambda }-\dot{\Sigma}_{\Lambda
}C^{2})H_{4}-6H_{6}C_{\Lambda }\dot{\Sigma}_{\Lambda },  \nonumber \\
&&...  \nonumber
\end{eqnarray}

Assuming that the relation (\ref{CL}) between the full and bare Green
functions is fulfilled during the flow, the scheme (\ref{Mod}) can be
obtained from equations (\ref{Polch}) by substituting
\begin{eqnarray}
V_{2}%
\begin{array}{c}
=%
\end{array}%
H_{2}\left( C_{\Lambda }/C_{0,\Lambda }\right) ^{2} &=&-\left( C_{\Lambda
}/C_{0,\Lambda }\right) \Sigma _{\Lambda }  \nonumber \\
V_{4}=H_{4}\left( C_{\Lambda }/C_{0,\Lambda }\right) ^{4}, &&...  \label{rep}
\end{eqnarray}%
For the choice (\ref{CL}) we also have $\dot{C}_{\Lambda }-\dot{\Sigma}%
_{\Lambda }C_{\Lambda }^{2}=\dot{C}_{0,\Lambda }/(1-C_{0,\Lambda }\Sigma
_{\Lambda })^{2}$, which allows to recover the equations for amputated
vertices, derived in Ref. \cite{SalmhoferBook}. However, as it was already
mentioned, general equations (\ref{Mod}) do not assume the validity of the
relation (\ref{CL}) during the flow; they are valid for any functional $%
C_{\Lambda }[\Sigma _{\Lambda }],$ which fulfills $C_{\Lambda =\Lambda
_{0}}=0$ in the beginning of the flow and the relation (\ref{CL}) in the end
of the flow.

The equation for the self-energy (\ref{Mod}) is similar to its form in 1PI
scheme (and coincides with this scheme for the choice (\ref{CL})). However,
the equations for higher-order vertices are organized differently: the pairs
of vertices are connected only by tree-like diagrams, while one- and higher
loop contributions are obtained by substituting results from higher- to
lower order vertices, which may lead to difficulties in the strong-coupling
regime. These equations also do not allow to see easily how the mean-field
results are reproduced.

\section{Wick-ordered schemes}

\subsection{General consideration and non-local schemes}

To overcome these difficulties, we following to Refs. \cite%
{SalmhoferBook,Salmhoferadj} consider the Wick ordered modification of Eqs. (%
\ref{Polch}). This scheme considers an expansion of the effective action in
the Wick-ordered monomials $\Omega _{D_{\Lambda }}(\Psi (x_{1})....\Psi
(x_{m}))$
\begin{eqnarray}
\mathcal{V}_{\Lambda } &=&\sum\limits_{m}W_{m}\Omega _{D_{\Lambda }}(\Psi
(x_{1})....\Psi (x_{m}))=e^{-\Delta _{D_{\Lambda }}}\widetilde{\mathcal{V}}%
_{\Lambda }  \label{relW} \\
\widetilde{\mathcal{V}}_{\Lambda } &=&\sum\limits_{m}W_{m}\Psi
(x_{1})....\Psi (x_{m})
\end{eqnarray}%
where the Wick-ordering propagator $D_{\Lambda }$ fulfills $D_{\Lambda =0}=0$%
. Performing Wick-ordering of equations (\ref{Eq}) using (\ref{relW}) and
assuming $W_{2}=0,$ we obtain%
\begin{eqnarray}
\dot{\Sigma}_{\Lambda } &=&-\left[ \frac{d(D_{\Lambda }+C_{\Lambda })}{%
d\Lambda }-\dot{\Sigma}_{\Lambda }C_{\Lambda }(2D_{\Lambda }+C_{\Lambda })%
\right] H_{4}  \label{RGEqGeneral} \\
&&+\sum_{m_{1},m_{2}\geq 4}H_{m_{1}}\circ (\dot{C}_{\Lambda }-\dot{\Sigma}%
_{\Lambda }C_{\Lambda }^{2})\circ D_{\Lambda }^{\frac{m_{1}+m_{2}}{2}%
-2}\circ H_{m_{2}}  \nonumber \\
\dot{H}_{m} &=&\left[ \frac{d(D_{\Lambda }+C_{\Lambda })}{d\Lambda }-\dot{%
\Sigma}_{\Lambda }C_{\Lambda }(2D_{\Lambda }+C_{\Lambda })\right]
H_{m+2}-mH_{m}\dot{\Sigma}_{\Lambda }C_{\Lambda }  \nonumber \\
&&-\sum_{m_{1},m_{2}\geq 4}H_{m_{1}}\circ (\dot{C}_{\Lambda }-\dot{\Sigma}%
_{\Lambda }C_{\Lambda }^{2})\circ D_{\Lambda }^{\frac{m_{1}+m_{2}-m}{2}%
-1}\circ H_{m_{2}}\ \ \ \ (m%
\begin{array}{c}
\geq%
\end{array}%
4)  \nonumber
\end{eqnarray}%
In the square brackets we have grouped together terms, corresponding to
generalised tadpole diagrams. Standard choice \cite{SalmhoferBook} is $\dot{D%
}_{\Lambda }=-\dot{C}_{\Lambda ,0}$. Let however choose $D$ such that
tadpole diagrams, including the contributions proportional to $\dot{\Sigma}$%
, cancel out. Then we have the following differential equation for $D:$%
\begin{equation}
\frac{d(D_{\Lambda }+C_{\Lambda })}{d\Lambda }-\dot{\Sigma}_{\Lambda
}C_{\Lambda }(2D_{\Lambda }+C_{\Lambda })=0  \label{DEq1}
\end{equation}%
The resulting hierarchy of equations is simplified and read%
\begin{eqnarray}
\dot{\Sigma}_{\Lambda } &=&\sum_{m_{1},m_{2}\geq 4}H_{m_{1}}\circ (\dot{C}%
_{\Lambda }-\dot{\Sigma}_{\Lambda }C_{\Lambda }^{2})\circ D_{\Lambda }^{%
\frac{m_{1}+m_{2}}{2}-2}\circ H_{m_{2}}  \label{RGEqRmTadpole} \\
\dot{H}_{m} &=&-mH_{m}\dot{\Sigma}_{\Lambda }C_{\Lambda }  \nonumber \\
&&-\sum_{m_{1},m_{2}\geq 4}H_{m_{1}}\circ (\dot{C}_{\Lambda }-\dot{\Sigma}%
_{\Lambda }C_{\Lambda }^{2})\circ D_{\Lambda }^{\frac{m_{1}+m_{2}-m}{2}%
-1}\circ H_{m_{2}}\ \ \ \ (m%
\begin{array}{c}
\geq%
\end{array}%
4)  \nonumber
\end{eqnarray}%
Similarly to Ref. \cite{Salmhoferadj}, to guarantee that $D_{\Lambda =0}=0$
one has to solve Eq. (\ref{DEq1}) in terms of corresponding condition at $%
\Lambda =0,$ instead of using initial condition at $\Lambda _{0}$. The
corresponding solution to the equation (\ref{DEq1}) has the form%
\begin{equation}
D_{\Lambda }{=}\,\,-\int_{0}^{\Lambda }d\Lambda ^{\prime }(\dot{C}_{\Lambda
^{\prime }}-\dot{\Sigma}_{\Lambda ^{\prime }}C_{\Lambda ^{\prime }}^{2})\exp
\left( 2\int_{\Lambda ^{\prime }}^{\Lambda }d\Lambda ^{\prime \prime
}C_{\Lambda ^{\prime \prime }}\dot{\Sigma}_{\Lambda ^{\prime \prime }}\right)
\label{EqDG}
\end{equation}%
In particular, for the choice of the propagator (\ref{CL}) we obtain
\begin{eqnarray}
D_{\Lambda } &=&\,\,\int_{0}^{\Lambda }d\Lambda ^{\prime }\dot{D}_{\Lambda
^{\prime },0}\exp \left( 2\int_{\Lambda ^{\prime }}^{\Lambda }d\Lambda
^{\prime \prime }C_{\Lambda ^{\prime \prime }}\dot{\Sigma}_{\Lambda ^{\prime
\prime }}\right)  \nonumber \\
&{=}&\frac{1}{(1-C_{0,\Lambda }\Sigma _{\Lambda })^{2}}\int_{0}^{\Lambda
}d\Lambda ^{\prime }\dot{C}_{0,\Lambda }^{\prec }\exp \left( 2\int_{\Lambda
^{\prime }}^{\Lambda }d\Lambda ^{\prime \prime }\frac{\Sigma _{\Lambda
^{\prime \prime }}\dot{C}_{0,\Lambda }^{\prec }}{1-C_{0,\Lambda ^{\prime
\prime }}\Sigma _{\Lambda ^{\prime \prime }}}\right)  \label{EqD}
\end{eqnarray}%
and%
\begin{eqnarray}
\dot{\Sigma}_{\Lambda } &=&\frac{1}{2}\sum\limits_{m_{1},m_{2}\geq
4}H_{m_{1}}\circ \frac{1}{(1-C_{0,\Lambda }\Sigma _{\Lambda })^{2}}\dot{C}%
_{0,\Lambda }^{\prec }\circ D_{\Lambda }^{\frac{m_{1}+m_{2}}{2}-2}\circ
H_{m_{2}}  \nonumber \\
\dot{H}_{m} &=&-mH_{m}C_{\Lambda }\dot{\Sigma}_{\Lambda }  \label{NewWick} \\
&&-\frac{1}{2}\sum\limits_{m_{1},m_{2}\geq 4}H_{m_{1}}\circ \frac{1}{%
(1-C_{0,\Lambda }\Sigma _{\Lambda })^{2}}\dot{C}_{0,\Lambda }^{\prec }\circ
D_{\Lambda }^{\frac{m_{1}+m_{2}-m}{2}-1}\circ H_{m_{2}}  \nonumber
\end{eqnarray}%
where $\dot{C}_{0,\Lambda }^{\prec }=-\dot{C}_{0,\Lambda }$ and $\dot{D}%
_{\Lambda ,0}=\dot{C}_{0,\Lambda }^{\prec }/(1-C_{\Lambda }^{0}\Sigma
_{\Lambda })^{2}$.

Let us see, how the mean-field solution is recovered from the equations (\ref%
{RGEqGeneral}) or (\ref{NewWick}). Since there are no tadpole diagrams for $%
d\Sigma _{\Lambda }/d\Lambda ,$ we have $d\Sigma _{\Lambda }^{\mathrm{MF}%
}/d\Lambda =0,$ i.e. $\Sigma =\Sigma ^{\mathrm{MF}}$ is constant in $\Lambda
$ and equal to its mean-field value, which is given by a (self-consistently
determined) sum of tadpole diagrams, absorbed into the definition of Wick
ordering$.$ Equation (\ref{EqDG}) is easily integrated and yields
\begin{equation}
D_{\Lambda }^{\mathrm{MF}}=C_{\Lambda =0}^{\mathrm{MF}}-C_{\Lambda }^{%
\mathrm{MF}}  \label{Dphys}
\end{equation}%
In case of fermions and moving Fermi surface (Re$\Sigma \neq 0$) this
function for smooth cutoff and intermediate $\Lambda $ has singularity at
both, physical and running Fermi surface. Since $\dot{D}_{\Lambda }^{\mathrm{%
MF}}=-\dot{C}_{\Lambda }^{\mathrm{MF}},$ the equation for $H_{4}$ can be
then also solved analytically to reproduce the standard random-phase
approximation (RPA) result%
\begin{equation}
H_{4}=\frac{H_{4}^{\mathrm{MF}}}{1-H_{4}^{\mathrm{MF}}\mathrm{Tr}(C_{\Lambda
}^{\mathrm{MF}})^{2}}.
\end{equation}

Returning to the treatment of the interaction beyond mean-filed theory, the
advantage of the equations (\ref{EqD}) and (\ref{NewWick}) is in their
simple form in the sharp cutoff limit. Assuming $\chi _{>,\Lambda }=\theta
(|\varepsilon _{\mathbf{k}}|-\Lambda ),$ $C_{0,\Lambda }=\chi _{>,\Lambda
}/(i\omega _{n}-\varepsilon _{\mathbf{k}}),$ we obtain%
\begin{equation}
D_{\Lambda }=\frac{\theta (\Lambda -|\varepsilon _{\mathbf{k}}|)}{i\omega
_{n}-\varepsilon _{\mathbf{k}}-\Sigma _{\Lambda ^{\prime }=|\varepsilon _{%
\mathbf{k}}|}}  \label{DL_sharp}
\end{equation}%
which requires the knowledge of $\Sigma _{\Lambda ^{\prime }}$ at $\Lambda
^{\prime }<\Lambda $, which has to be determined self-consistently and the
single-scale propagator%
\begin{equation}
F_{\Lambda }=\frac{\delta (\Lambda -|\varepsilon _{\mathbf{k}}|)}{i\omega
_{n}-\varepsilon _{\mathbf{k}}-\Sigma _{\Lambda }},  \label{SL_sharp}
\end{equation}%
Both propagators have simpler form, than those in Ref. \cite{Salmhoferadj},
where additional integration over the cutoff parameter is still present even
for the sharp cutoff (see Appendix).

The equations (\ref{RGEqGeneral}) can be somewhat generalized by considering
reamputation of vertices $H_{\Lambda }^{(n)}=\widetilde{H}_{\Lambda
}^{(n)}(S_{\Lambda })^{n}$ with respect to the amputation by full Green
functions $C_{\Lambda }$ (corresponding to $S_{\Lambda }=1$),
\begin{eqnarray}
S_{\Lambda }^{-2}\dot{\Sigma}_{\Lambda } &=&-\left[ \frac{d\widetilde{D}%
_{\Lambda }}{d\Lambda }-2\widetilde{D}_{\Lambda }(\dot{\Sigma}_{\Lambda
}C_{\Lambda }+\dot{S}_{\Lambda }S_{\Lambda }^{-1})+S_{\Lambda }^{2}(\dot{C}%
_{\Lambda }-\dot{\Sigma}_{\Lambda }C_{\Lambda }^{2})\right] \circ \widetilde{%
H}_{4}  \nonumber \\
&&+\sum\limits_{m_{1},m_{2}\geq 4}\widetilde{H}_{m_{1}}\circ S_{\Lambda
}^{2}(\dot{C}_{\Lambda }-\dot{\Sigma}_{\Lambda }C_{\Lambda }^{2})\circ
\widetilde{D}_{\Lambda }^{\frac{m_{1}+m_{2}}{2}-2}\circ \widetilde{H}_{m_{2}}
\nonumber \\
\dot{\widetilde{H}}_{m} &=&\left[ \frac{d\widetilde{D}_{\Lambda }}{d\Lambda }%
-2\widetilde{D}_{\Lambda }(\dot{\Sigma}_{\Lambda }C_{\Lambda }+\dot{S}%
_{\Lambda }S_{\Lambda }^{-1})+S_{\Lambda }^{2}(\dot{C}_{\Lambda }-\dot{\Sigma%
}_{\Lambda }C_{\Lambda }^{2})\right] \circ \widetilde{H}_{m+2}  \nonumber \\
&&-\sum\limits_{m_{1},m_{2}\geq 4}\widetilde{H}_{m_{1}}\circ S_{\Lambda
}^{2}(\dot{C}_{\Lambda }-\dot{\Sigma}C_{\Lambda }^{2})\circ \widetilde{D}%
_{\Lambda }^{\frac{m_{1}+m_{2}-m}{2}-1}\circ \widetilde{H}_{m_{2}}  \nonumber
\\
&&-m\widetilde{H}_{m}(C_{\Lambda }\dot{\Sigma}_{\Lambda }+\dot{S}_{\Lambda
}S_{\Lambda }^{-1})  \label{RGEqGeneralS}
\end{eqnarray}%
where $\widetilde{D}_{\Lambda }=D_{\Lambda }S_{\Lambda }^{2}$. Equations (%
\ref{RGEqGeneralS}) allow to eastablish connection with the dynamic
adjustment propagator scheme of Ref. \cite{Salmhoferadj}. The explicit
relation between the equations (\ref{RGEqGeneralS}) and those of Ref. \cite%
{Salmhoferadj} is discussed in the Appendix and involves passing from the
function $C_{\Lambda }$ to a different function $D_{\Lambda }^{\mathrm{M}}$
with the use of equation (\ref{rel}).

If, similarly to Ref. \cite{Salmhoferadj}, we require $C_{\Lambda }\dot{%
\Sigma}_{\Lambda }+\dot{S}_{\Lambda }S_{\Lambda }^{-1}=0,$ than equations (%
\ref{RGEqGeneralS}) take simpler form due to cancellation of some terms in
this specially chosen amputation of vertices. This way, equations (\ref%
{RGEqGeneralS}) allow also to eastablish simpler view on the result (\ref%
{EqDG}). Indeed, let assume that $\dot{\Sigma}_{\Lambda }C_{\Lambda }+\dot{S}%
_{\Lambda }S_{\Lambda }^{-1}=0$ for some $S.$ Then we have
\begin{equation}
S_{\Lambda }=\exp \left( \int_{\Lambda }^{\Lambda _{0}}\dot{\Sigma}_{\Lambda
^{\prime }}C_{\Lambda ^{\prime }}d\Lambda ^{\prime }\right)  \label{Sntp}
\end{equation}%
The condition of vanishing generalized tadpole diagrams yields $\frac{d%
\widetilde{D}_{\Lambda }}{d\Lambda }=-S_{\Lambda }^{2}(\dot{C}_{\Lambda }-%
\dot{\Sigma}_{\Lambda }C_{\Lambda }^{2})$, which leads us to equation (\ref%
{EqDG}). Although the condition (\ref{EqDG}) generally does not yield Wick
propagators, which derivative coincides with the single-scale propagator $%
\dot{\Sigma}_{\Lambda }C_{\Lambda }^{2}-\dot{C}_{\Lambda }$, this is
achieved in the amputated scheme with the amputation factor $S_{\Lambda }$
given by the equation (\ref{Sntp}). In general, analytical or numerical
evaluation of the integral in this equation can be, however, rather
complicated.

\subsection{Local schemes}

In practical applications, the schemes, which are local in $\Lambda $, may
have some advantage. Simplest choice of the propagators%
\begin{equation}
D_{\Lambda }=\frac{\chi _{<,\Lambda }}{C_{0}^{-1}-\Sigma _{\Lambda }}%
;C_{\Lambda }=\frac{\chi _{>,\Lambda }}{C_{0}^{-1}-\Sigma _{\Lambda }}
\label{CD1}
\end{equation}%
suffers from the problem, discussed in Ref. \cite{Salmhoferadj}, namely it
yields remaining tadpole terms in $\partial _{\Lambda }H_{m}$, proportional
to $\left( D_{\Lambda }^{2}\partial _{\Lambda }\Sigma _{\Lambda }\right)
\circ H_{m+2}$, which have potential infrared divergencies due to square of
the propagator $D_{\Lambda }$. To avoid this problem, we choose the
propagator $C_{\Lambda }$ according to the equation (\ref{CL}) and use
similar expression for $D_{\Lambda }$, including the mean-field self-energy
into the bare propagator and choosing non-interacting part of the action $%
C_{0,\Lambda }=\chi _{>,\Lambda }\overline{C}_{0,\Lambda }$,
\begin{equation}
D_{\Lambda }=\frac{\chi _{<,\Lambda }}{\overline{C}_{0,\Lambda }^{-1}-\Sigma
^{\mathrm{MF}}-\chi _{>,\Lambda }\Sigma _{\Lambda }};C_{\Lambda }=\frac{\chi
_{>,\Lambda }}{\overline{C}_{0,\Lambda }^{-1}-\Sigma ^{\mathrm{MF}}-\chi
_{>,\Lambda }\Sigma _{\Lambda }},  \label{CD2}
\end{equation}%
to obtain from the equations (\ref{RGEqGeneral}) the local flow equations%
\begin{eqnarray}
\dot{\Sigma}_{\Lambda } &=&-T_{\Lambda }\circ H_{4}  \label{E2} \\
&&-\sum\limits_{m_{1},m_{2}\geq 4}H_{m_{1}}\circ F_{\Lambda }\circ
D_{\Lambda }^{\frac{m_{1}+m_{2}}{2}-2}\circ H_{m_{2}},  \nonumber \\
\dot{H}_{m} &=&T_{\Lambda }\circ H_{m+2}-mH_{m}C_{\Lambda }\dot{\Sigma}%
_{\Lambda }  \nonumber \\
&&+\sum\limits_{m_{1},m_{2}\geq 4}H_{m_{1}}\circ F_{\Lambda }\circ
D_{\Lambda }^{\frac{m_{1}+m_{2}-m}{2}-1}\circ H_{m_{2}}.  \nonumber
\end{eqnarray}%
where
\begin{eqnarray}
T_{\Lambda } &=&\frac{r\overline{C}_{0,\Lambda }^{-2}\dot{\overline{C}}%
_{0,\Lambda }\chi _{>,\Lambda }+r\Sigma _{\Lambda }\dot{\chi}_{>,\Lambda
}-\chi _{<,\Lambda }\chi _{>,\Lambda }\dot{\Sigma}_{\Lambda }}{(\overline{C}%
_{0,\Lambda }^{-1}-\Sigma ^{\mathrm{MF}}-\chi _{>,\Lambda }\Sigma _{\Lambda
})^{2}}  \label{TF2} \\
F_{\Lambda } &=&-\frac{\overline{C}_{0,\Lambda }^{-2}\dot{\overline{C}}%
_{0,\Lambda }\chi _{>,\Lambda }+(\overline{C}_{0,\Lambda }^{-1}-\Sigma ^{%
\mathrm{MF}})\dot{\chi}_{>,\Lambda }}{(\overline{C}_{0,\Lambda }^{-1}-\Sigma
^{\mathrm{MF}}-\chi _{>,\Lambda }\Sigma _{\Lambda })^{2}}  \nonumber
\end{eqnarray}%
and we have assumed $\chi _{>,\Lambda }+\chi _{<,\Lambda }=r$. Due to
special choice of propagators, the solution to the equations (\ref{E2}) has
natural physical interpretation even at the intermediate stages of the flow,
since it corresponds to the flow of the functional with the bare propagator $%
\chi _{>,\Lambda }/(C_{0,\Lambda }^{-1}-\Sigma ^{\mathrm{MF}})$ after the
Wick ordering with the propagator $D_{\Lambda }$. The equations (\ref{E2})
have to be solved with the initial condition $\Sigma _{\Lambda }=0$, since
the constant (mean-field) initial part of $\Sigma _{\Lambda }$ enters
propagators explicitly through $\Sigma ^{\mathrm{MF}}.$

The scheme (\ref{CD2})-(\ref{TF2}) may be useful for the realization of the
interaction \cite{HonerU} and temperature- \cite{Honer,KataninKampf,Katanin1}
flow in the Wick-ordered scheme. Indeed, the choice $\chi _{>,\Lambda
}=\Lambda ^{1/2}$, $\chi _{<,\Lambda }=1-\Lambda ^{1/2},$ and $\overline{C}%
_{0,\Lambda }=C_{0}$ in the equation (\ref{CD1}) is analogous to the
interaction flow \cite{HonerU}, while the choice $\Lambda =T,\ \chi
_{>,T}=T^{-1/2}$, $r=T_{0}^{-1/2}$, and $\overline{C}_{0,T}^{-1}=(i\omega
_{n}-\varepsilon _{\mathbf{k}})/T$, where $T$ is the temperature and $T_{0}$
is the final temperature of the flow, is analogous to the temperature-flow
in the 1PI approach \cite{Honer,KataninKampf,Katanin1}. The disadvantage of
the latter scheme in Wick-ordered approach is that similarly to the 1PI flow
with counterterms \cite{1PIcounter} it requires the final temperature of the
flow as an input, and therefore does not allow to obtain the whole
temperature evolution of the system for the same input parameters.

Within similar truncations of the hierarchy, the momentum-, frequency, or
interaction flow in the Wick-ordered scheme with the choice of propagators (%
\ref{CD2}) can be already superior to the 1PI scheme, since the latter
contains tadpole terms of the order $O(H_{m+2})$ in the weak-coupling
regime, which are needed to generate multiple-loop contributions from fRG
hierarchy. On contrary, the Wick-ordered scheme contains these contributions
explicitly, while the \textquotedblleft dangerous\textquotedblright\ tadpole
contributions have smaller value; the initial (mean-field) self-energy is
also already included in both propagators $C_{\Lambda }$ and $D_{\Lambda }.$
Specifically, in the weak-coupling limit in the scheme (\ref{E2}) we obtain $%
\Sigma _{\Lambda }\sim \dot{\Sigma}_{\Lambda }=O(H_{4}^{2})$, yielding
therefore the correction to $\dot{\Sigma}_{\Lambda }$ from the remaining
tadpole diagrams of the order $O(H_{4}^{3})$ and the corresponding
correction to the vertices $H_{m\geq 4}$ of the order $O(H_{4}^{2}H_{m+2})$.

Smaller value of the tadpole terms in Wick-ordered scheme and explicit
presence of multi-loop contributions also implies possibility of having less
fine parametrization of vertices, required for the solution of the equations
(\ref{E2}) and providing the same accuracy, as in 1PI scheme. Indeed, the
effort of extra evaluation of each loop in Wick ordered scheme scales as $%
n_{p}n_{o}$ where $n_{p}$ is the number of integration points in
momentum-frequency space and $n_{o}$ is the number of internal (e.g. spin,
orbital, etc.) degrees of freedom. At the same time, finer parametrization
of $m$-point vertex in 1PI scheme increases computational effort by $%
(n_{1}/n_{2})^{m-1}$ where $n_{1}$ and $n_{2}<n_{1}$ are the number of
patches in momentum-frequency space for the 1PI and Wick-ordered equations,
respectively. For $n_{1}\sim n_{p}/10,$ $n_{2}\sim 10$ and for not very
large number of internal degrees of freedom, the increase of the effort in
1PI scheme $(n_{1}/n_{2})^{m-1},$ required to reproduce accurately
multi-loop contributions, can easily exceed $n_{p}n_{o}$ even for the $m=4$%
-point vertex due to extra factor $n_{p}^{2}/(10^{6}n_{o}),$ which is rather
large in the practical applications, because of large $n_{p}\sim 10^{4}$.

Application of the full cancellation of the tadpole diagrams, given by the
equation (\ref{EqD}) or dynamical adjusting scheme of Ref. \cite%
{Salmhoferadj} may further improve applicability of the Wick-ordered
equations, but is considerably more complicated because of the necessity to
fulfill the self-consistency condition $D_{\Lambda =0}^{\mathrm{phys}}=0$
(or $R_{\Lambda =0}=0$), which results in the requirement to know the
self-energy $\Sigma _{\Lambda ^{\prime }}$ at the stages of the flow, which
are later than the current one. Yet, even in this case one can search for
easier integrable functions by adjusting $C_{\Lambda }$ in equations (\ref%
{RGEqGeneral}), (\ref{RGEqGeneralS}).

\subsection{Toy model}

To gain insight into the applicability of the described approaches, we
consider the toy $\varphi ^{4}$-like model\cite{Meden}%
\begin{equation}
S_{0}=\frac{a}{2}\varphi ^{2};\ \ \ \mathcal{V}=\frac{b}{4!}\varphi ^{4}
\label{fimodel}
\end{equation}%
In the following we put $a=1/\chi _{>,\Lambda }$ with $\chi _{>,\Lambda
}=1-\Lambda $ and use $\Lambda $ as a scaling parameter, which changes from
one (bare model) to zero. In the truncation $H_{6}=0$ the flow equations (%
\ref{DEq1}), (\ref{NewWick}) take the form 
\begin{eqnarray}
\dot{\Sigma}_{\Lambda } &=&-\frac{1}{2}b_{\Lambda }^{2}F_{\Lambda
}D_{\Lambda }^{2}  \nonumber \\
\dot{b}_{\Lambda } &=&\ \ \ \ \ 3b_{\Lambda }^{2}F_{\Lambda }D_{\Lambda
}-4b_{\Lambda }C_{\Lambda }\dot{\Sigma}_{\Lambda }  \label{tadmodel} \\
\dot{D}_{\Lambda } &=&F_{\Lambda }+2\dot{\Sigma}_{\Lambda }C_{\Lambda
}D_{\Lambda }  \nonumber
\end{eqnarray}%
where we choose, according to the equation (\ref{CD2}), 
\begin{equation}
C_{\Lambda }=\frac{1-\Lambda }{1-\Sigma ^{\mathrm{MF}}-(1-\Lambda )\Sigma
_{\Lambda }},  \label{Ctadmodel}
\end{equation}%
such that%
\begin{equation}
F_{\Lambda }=-\dot{C}_{\Lambda }+\dot{\Sigma}_{\Lambda }C_{\Lambda }^{2}=%
\frac{1-\Sigma ^{\mathrm{MF}}}{(1-\Sigma ^{\mathrm{MF}}-(1-\Lambda )\Sigma
_{\Lambda })^{2}};  \label{Ftadmodel}
\end{equation}%
the initial conditions $\Sigma ^{\mathrm{MF}}=\frac{1}{2}(1-\sqrt{1+2b}),$ $%
\Sigma _{\Lambda =1}=0,~b_{\Lambda =1}=b$ and $D_{\Lambda =0}=0$. In
numerical calculations, we adjust the initial value $D_{\Lambda =1}$ to
achieve vanishing of the propagator in the end of the flow.

On the other hand, the local equations (\ref{CD2}), (\ref{E2}) yield 
\begin{eqnarray}
\dot{\Sigma}_{\Lambda } &=&-\frac{1}{2}b_{\Lambda }T_{\Lambda }\ \ -\frac{1}{%
2}b_{\Lambda }^{2}F_{\Lambda }D_{\Lambda }^{2}  \label{RGfi} \\
\dot{b}_{\Lambda } &=&-4b_{\Lambda }C_{\Lambda }\dot{\Sigma}_{\Lambda
}+3b_{\Lambda }^{2}F_{\Lambda }D_{\Lambda }  \nonumber
\end{eqnarray}%
with 
\begin{eqnarray}
D_{\Lambda } &=&\frac{\Lambda }{1-\Sigma ^{\mathrm{MF}}-(1-\Lambda )\Sigma
_{\Lambda }}  \nonumber \\
T_{\Lambda } &=&\frac{\Sigma _{\Lambda }-\Lambda (1-\Lambda )\dot{\Sigma}%
_{\Lambda }}{(1-\Sigma ^{\mathrm{MF}}-(1-\Lambda )\Sigma _{\Lambda })^{2}}
\end{eqnarray}

The results of the solution of equations (\ref{tadmodel}) and (\ref{RGfi})
for the self-energy and coupling constant in the end of the flow for
interaction strength $b=1,2,3$ are presented in the Table. For comparison,
we also present the result of 1PI approach \cite{Meden} and the local
scheme, suggested in Ref. \cite{Salmhoferadj}, see also the equations (\ref%
{SalmhofLocal}) of the Appendix; we find numerically, that using the choice (%
\ref{CD1}) of propagators in the equations (\ref{RGfi}) yields the same
results, as the latter scheme. Note that, as discussed in previous Section,
the local scheme based on the choice (\ref{CD1}), as well as the local
scheme of Ref. \cite{Salmhoferadj}, are actually not suitable for treatment
of many-body systems because of possible divergencies in the flow equations.
The non-local (adjusting) scheme of Ref. \cite{Salmhoferadj} (see also
equations (\ref{SalmhofAdjust}) of the Appendix) is found to yield
numerically very close results to the equations (\ref{tadmodel})-(\ref%
{Ftadmodel}).

One can see from the Table, that for the model (\ref{fimodel}) all
considered Wick-ordered schemes yield improvement in comparison to 1PI
scheme; the equations (\ref{RGfi}) improve the results for vertices of the
local approach of Ref. \cite{Salmhoferadj}, the self-energies of the two
approaches are practically identical. Surprisingly, being computationally
more expensive and yielding slight improvement of 1PI scheme, the
approaches, fully excluding tadpole diagrams, yield worse results for the
considering toy model, than the local schemes. Therefore, the adjustment of
the Wick propagator to exclude fully tadpole diagrams does not generally
implies smaller trucation errors. This fact can be attributed to the
property of the Wick-ordered equations in general amputation scheme (\ref%
{RGEqGeneralS}) to have derivative of the Wick propagator non-equal to the
single-scale propagator. Therefore, some remaining tadpole terms seem
necessary to compensate this difference. Note that choosing special
amputation of the vertices, which provide equality of the derivative of the
Wick propagator to the single-scale propagator, such as given by the Eq. (%
\ref{Sntp}) or considered in Ref. \cite{Salmhoferadj}, does not yield
further improvement of the truncation, since the amputation itself does not
change physically observable quantities. For more complicated problems, one
may also expect that the schemes, which are local in the cutoff parameter,
but do not exclude fully tadpole diagrams (such as the scheme (\ref{E2}))
are more preferable. \medskip\ 

\begin{tabular}{||l||l|l|l||l|l|l||}
\hline\hline
Method & \multicolumn{3}{c||}{$-\Sigma $} & \multicolumn{3}{c||}{$b_{R}$} \\ 
\hline
\multicolumn{1}{||r||}{$b=$} & $1$ & $2$ & $3$ & $1$ & $2$ & $3$ \\ 
\hline\hline
1PI, Ref. \cite{Meden} & 0.320398 & 0.519824 & 0.677443 & 0.513382 & 0.809275
& 1.05686 \\ \hline
Eqs. (\ref{RGEqRmTadpole}), Eqs. (\ref{tadmodel}) & 0.323881 & 0.527393 & 
0.687805 & 0.530763 & 0.855922 & 1.13321 \\ \hline
local, Ref.\cite{Salmhoferadj} & 0.330197 & 0.546350 & 0.720883 & 0.535013 & 
0.874050 & 1.17181 \\ \hline
Eqs.(\ref{E2});Eqs.(\ref{RGfi}) & 0.329972 & 0.545415 & 0.719051 & 0.542277
& 0.899275 & 1.21913 \\ \hline
Exact & 0.332425 & 0.550557 & 0.726505 & 0.607899 & 1.055430 & 1.46469 \\ 
\hline\hline
\end{tabular}%
\smallskip

Table. The self-energy and connected fully amputated 4-point vertex $b_{R}$
of the model (\ref{fimodel}) obtained by different methods for values of the
coupling constant $b=1,2,3$.

\section{Summary}

In Summary, we have considered the renormalization-group schemes, which
allow to treat self-energy effects in a non-perturbative way and do not rely
on using one-particle irreducible functionals. In the Wick-ordered scheme
the choice of the propagator, with respect to which the Wick-ordering is
performed, allowing to remove partly or fully tadpole diagrams from
renormalization-group equations, has some advantage over 1PI formalism.
While full removing of tadpole terms yields the differential equation with
\textquotedblleft finite\textquotedblright\ instead of initial condition,
and the results, which are similar to the earlier proposed approach \cite%
{Salmhoferadj}, removing partly tadpole terms we have obtained the
differential equations, which are local with respect to the cutoff parameter
and avoid potential divergencies in the flow equations. We have also argued
that in general the suggested local approach may be superior to 1PI
formalism, due to smaller contribution of the tadpole diagrams, which
implies that in general it requires less fine parametrization of the
vertices, that the 1PI approach. Further investigations of the necessity to
keep some tadpole contributions in the Wick-ordering hierarchy and
comparison of 1PI and Wick-ordered schemes for more complicated models would
be of certain interest.

\textit{Acknowledgements}. The author is grateful to M. Salmhofer for
discussions on the properties of Wick-ordered approach and hospitability
during the visit to the Institute of Theoretical Physics (Heidelberg), and
also to Max-Planck Society for partial financial support within the
Partnership Program.

\section*{Appendix. Relation to the dynamic adjustment scheme of Ref. 
\protect\cite{Salmhoferadj}.}

In this Appendix we consider relation of the approach of present paper to
that of Ref. \cite{Salmhoferadj}. To relate the equations (\ref{RGEqGeneral}%
) to those of Ref. \cite{Salmhoferadj}, we put 
\begin{equation}
C_{\Lambda }S^{2}=\mathcal{C}_{\Lambda }-D_{\Lambda }^{\mathrm{M}}
\label{rel}
\end{equation}%
with some function $D_{\Lambda }^{M},$ which satisfies $D_{\Lambda =\Lambda
_{0}}^{M}=\mathcal{C}_{\Lambda =\Lambda _{0}}$. Choosing $\mathcal{C}%
=S^{2}/(C_{0}^{-1}-\Sigma _{\Lambda })$, we obtain%
\begin{eqnarray}
S^{2}(\dot{C}_{\Lambda }-\dot{\Sigma}_{\Lambda }C_{\Lambda }^{2}) &=&S^{2}(%
\frac{d}{d\Lambda }(\mathcal{C}/S^{2})-\dot{\Sigma}_{\Lambda }\mathcal{C}%
^{2}/S^{4})-\frac{dD_{\Lambda }^{\mathrm{M}}}{d\Lambda } \\
&&+2(\dot{\Sigma}_{\Lambda }/S^{2}\mathcal{C}+S_{\Lambda }^{-1}\dot{S}%
_{\Lambda })D_{\Lambda }^{\mathrm{M}}-\dot{\Sigma}_{\Lambda }/S^{2}\left(
D_{\Lambda }^{\mathrm{M}}\right) ^{2}  \nonumber \\
&=&-\frac{dD_{\Lambda }^{\mathrm{M}}}{d\Lambda }+2S_{2\Lambda }^{-1}\dot{S}%
_{2\Lambda }D_{\Lambda }^{\mathrm{M}}-\dot{\Sigma}_{\Lambda }S^{-2}\left(
D_{\Lambda }^{\mathrm{M}}\right) ^{2}  \nonumber
\end{eqnarray}%
where in the last line we have assumed $S=S_{1}S_{2}$ and have chosen $%
S_{1}=C_{0}^{-1}-\Sigma _{\Lambda }$. This choice of $S_{1}$ implies that
the remaining factor $S_{2}$ represents the reamputation of the vertices,
amputated by $C_{\Lambda }(C_{0}^{-1}-\Sigma _{\Lambda })$. With this choice
we also have 
\[
S^{-2}\dot{\Sigma}_{\Lambda }(\mathcal{C}-D_{\Lambda }^{\mathrm{M}})+\dot{S}%
_{\Lambda }S_{\Lambda }^{-1}=-\dot{\Sigma}_{\Lambda }S^{-2}D_{\Lambda }^{%
\mathrm{M}}+\dot{S}_{2\Lambda }/S_{2\Lambda } 
\]%
and 
\[
S^{-2}\dot{\Sigma}_{\Lambda }=S_{2\Lambda }^{-2}(\partial _{\Lambda }\Sigma
_{\Lambda })/(C_{0}^{-1}-\Sigma _{\Lambda })^{2}=S_{2\Lambda }^{-2}\partial
_{\Lambda }(1/(C_{0}^{-1}-\Sigma _{\Lambda }))=S_{2\Lambda }^{-2}\partial
_{\Lambda }G_{\Lambda } 
\]%
Introducing $\mathcal{Q}_{\Lambda }=S_{2\Lambda }^{-2}\partial _{\Lambda
}G_{\Lambda },$ we obtain%
\begin{eqnarray}
\mathcal{Q}_{\Lambda } &=&-\left[ \frac{d}{d\Lambda }(\widetilde{D}_{\Lambda
}-D_{\Lambda }^{\mathrm{M}})+2D_{\Lambda }(D_{\Lambda }^{\mathrm{M}}\mathcal{%
Q}_{\Lambda }-\dot{S}_{2\Lambda }S_{2\Lambda }^{-1})\right.  \nonumber \\
&&\left. -(D_{\Lambda }^{M})^{2}\mathcal{Q}_{\Lambda }+2S_{2\Lambda }^{-1}%
\dot{S}_{2\Lambda }D_{\Lambda }^{\mathrm{M}}\right] \widetilde{H}_{4} 
\nonumber \\
&&-\sum\limits_{m_{1},m_{2}\geq 4}\widetilde{H}_{m_{1}}\circ (\frac{%
dD_{\Lambda }^{\mathrm{M}}}{d\Lambda }+\mathcal{Q}_{\Lambda }(D_{\Lambda }^{%
\mathrm{M}})^{2}-2S_{2\Lambda }^{-1}\dot{S}_{2\Lambda }D_{\Lambda }^{\mathrm{%
M}})\circ D_{\Lambda }^{\frac{m_{1}+m_{2}}{2}-2}\circ \widetilde{H}_{m_{2}} 
\nonumber \\
\dot{\widetilde{H}}_{m} &=&\left[ \frac{d}{d\Lambda }(\widetilde{D}_{\Lambda
}-D_{\Lambda }^{\mathrm{M}})+2D_{\Lambda }(\mathcal{Q}_{\Lambda }D_{\Lambda
}^{\mathrm{M}}-\dot{S}_{2\Lambda }S_{2\Lambda }^{-1})\right.  \nonumber \\
&&\left. -(D_{\Lambda }^{\mathrm{M}})^{2}\mathcal{Q}_{\Lambda }+2S_{2\Lambda
}^{-1}\dot{S}_{2\Lambda }D_{\Lambda }^{\mathrm{M}}\right] \widetilde{H}_{m+2}
\nonumber \\
&&+\sum\limits_{m_{1},m_{2}\geq 4}\widetilde{H}_{m_{1}}\circ (\frac{%
dD_{\Lambda }^{\mathrm{M}}}{d\Lambda }+\mathcal{Q}_{\Lambda }(D_{\Lambda }^{%
\mathrm{M}})^{2}-2S_{2\Lambda }^{-1}\dot{S}_{2\Lambda }D_{\Lambda }^{\mathrm{%
M}})\circ D_{\Lambda }^{\frac{m_{1}+m_{2}-m}{2}-1}\circ \widetilde{H}_{m_{2}}
\nonumber \\
&&+m\widetilde{H}_{m}(D_{\Lambda }^{\mathrm{M}}\mathcal{Q}_{\Lambda }-\dot{S}%
_{2\Lambda }/S_{2\Lambda })  \label{MS}
\end{eqnarray}

For $D_{\Lambda }^{\mathrm{M}}\mathcal{Q}_{\Lambda }-\dot{S}_{2\Lambda
}/S_{2\Lambda }=0$ we obtain equations 
\begin{eqnarray}
\mathcal{Q}_{\Lambda } &=&-\left[ \frac{d}{d\Lambda }(\widetilde{D}_{\Lambda
}-D_{\Lambda }^{\mathrm{M}})+(D_{\Lambda }^{M})^{2}\mathcal{Q}_{\Lambda }%
\right] \widetilde{H}_{4}  \nonumber \\
&&-\sum\limits_{m_{1},m_{2}\geq 4}\widetilde{H}_{m_{1}}\circ (\frac{%
dD_{\Lambda }^{\mathrm{M}}}{d\Lambda }-\mathcal{Q}_{\Lambda }(D_{\Lambda }^{%
\mathrm{M}})^{2})\circ D_{\Lambda }^{\frac{m_{1}+m_{2}}{2}-2}\circ 
\widetilde{H}_{m_{2}}  \nonumber \\
\dot{\widetilde{H}}_{m} &=&\left[ \frac{d}{d\Lambda }(\widetilde{D}_{\Lambda
}-D_{\Lambda }^{\mathrm{M}})+(D_{\Lambda }^{\mathrm{M}})^{2}\mathcal{Q}%
_{\Lambda }\right] \widetilde{H}_{m+2}  \nonumber \\
&&+\sum\limits_{m_{1},m_{2}\geq 4}\widetilde{H}_{m_{1}}\circ (\frac{%
dD_{\Lambda }^{\mathrm{M}}}{d\Lambda }-\mathcal{Q}_{\Lambda }(D_{\Lambda }^{%
\mathrm{M}})^{2})\circ D_{\Lambda }^{\frac{m_{1}+m_{2}-m}{2}-1}\circ 
\widetilde{H}_{m_{2}}
\end{eqnarray}%
which coincide with the equations of Ref. \cite{Salmhoferadj}. Choosing the
propagators as in Ref. \cite{Salmhoferadj}, $D_{M}^{\Lambda }=\chi
_{<,\Lambda }/\mathcal{A}_{\Lambda },$ $S_{2\Lambda }=1/\mathcal{A}_{\Lambda
},$ $\frac{d\mathcal{A}_{\Lambda }}{d\Lambda }=-\chi _{<,\Lambda }\mathcal{Q}%
_{\Lambda },$ we obtain 
\begin{eqnarray}
\mathcal{Q}_{\Lambda } &=&-\sum\limits_{m_{1},m_{2}\geq 4}\widetilde{H}%
_{m_{1}}\circ \frac{d\widetilde{D}_{\Lambda }}{d\Lambda }\circ \widetilde{D}%
_{\Lambda }^{\frac{m_{1}+m_{2}}{2}-2}\circ \widetilde{H}_{m_{2}},  \nonumber
\\
\dot{\widetilde{H}}_{m} &=&\sum\limits_{m_{1},m_{2}\geq 4}\widetilde{H}%
_{m_{1}}\circ \frac{d\widetilde{D}_{\Lambda }}{d\Lambda }\circ \widetilde{D}%
_{\Lambda }^{\frac{m_{1}+m_{2}-m}{2}-1}\circ \widetilde{H}_{m_{2}},
\label{SalmhofAdjust}
\end{eqnarray}%
where the single-scale and Wick propagators are given by%
\begin{eqnarray}
\ \widetilde{D}_{\Lambda } &=&\int_{0}^{\Lambda }d\Lambda ^{\prime
}F_{\Lambda ^{\prime }},\ \ F_{\Lambda }=\frac{\dot{\chi}_{<,\Lambda }}{%
\mathcal{A}_{\Lambda }}  \nonumber \\
\mathcal{A}_{\Lambda } &=&\frac{1}{C_{0}^{-1}-\Sigma _{\Lambda
_{0}}+\int_{\Lambda }^{\Lambda _{0}}d\Lambda ^{\prime \prime }\chi
_{<,\Lambda ^{\prime \prime }}\mathcal{Q}_{\Lambda ^{\prime \prime }}}.
\label{EqSalmhof}
\end{eqnarray}%
The relations (\ref{EqSalmhof}) are not easily simplified further for the
sharp cutoff because of the involved relation between $\mathcal{Q}_{\Lambda
} $ and $\Sigma _{\Lambda },$ $\mathcal{Q}_{\Lambda }=[\mathcal{A}_{\Lambda
}/(C_{0}^{-1}-\Sigma _{\Lambda })]^{2}\dot{\Sigma}_{\Lambda }$.

In the local scheme $\widetilde{D}_{\Lambda }=D_{\Lambda }^{\mathrm{M}}$,
also considered in Ref. \cite{Salmhoferadj}, we have%
\begin{eqnarray}
\mathcal{Q}_{\Lambda } &=&-\left( \frac{\chi _{<,\Lambda }}{\mathcal{A}%
_{\Lambda }}\right) ^{2}\mathcal{Q}_{\Lambda }\circ \widetilde{H}_{4} 
\nonumber \\
&&-\sum\limits_{m_{1},m_{2}\geq 4}\widetilde{H}_{m_{1}}\circ \left( \frac{%
\dot{\chi}_{<,\Lambda }}{\mathcal{A}_{\Lambda }}\right) \circ \left( \frac{%
\chi _{<,\Lambda }}{\mathcal{A}_{\Lambda }}\right) ^{\frac{m_{1}+m_{2}}{2}%
-2}\circ \widetilde{H}_{m_{2}}  \nonumber \\
\dot{\widetilde{H}}_{m} &=&\left( \frac{\chi _{<,\Lambda }}{\mathcal{A}%
_{\Lambda }}\right) ^{2}\mathcal{Q}_{\Lambda }\circ \widetilde{H}_{m+2} 
\nonumber \\
&&+\sum\limits_{m_{1},m_{2}\geq 4}\widetilde{H}_{m_{1}}\circ \left( \frac{%
\dot{\chi}_{<,\Lambda }}{\mathcal{A}_{\Lambda }}\right) \circ \left( \frac{%
\chi _{<,\Lambda }}{\mathcal{A}_{\Lambda }}\right) ^{\frac{m_{1}+m_{2}-m}{2}%
-1}\circ \widetilde{H}_{m_{2}}  \label{SalmhofLocal}
\end{eqnarray}%
As well as the scheme (\ref{CD1}), the latter equations may suffer for
general many-body problems from divergencies of the convolution of $%
\widetilde{H}_{m\geq 4}$ with the square of propagator $\chi _{<,\Lambda }/%
\mathcal{A}_{\Lambda }$ and $\mathcal{Q}_{\Lambda }$, which are not cut in
momentum space.

\end{document}